\documentclass[journal,12pt,onecolumn,draftclsnofoot,]{IEEEtran}
\usepackage{amssymb}
\usepackage{amsmath}

\usepackage{array}

\usepackage[dvipsnames]{xcolor}	

\usepackage{pgfplots}
\usepackage{algorithm, algpseudocode}
\usepackage{enumerate}
\usepackage{graphicx}
\usepackage{bm}

\usepackage[nospace]{cite}
\renewcommand{\baselinestretch}{0.94}

\begin{document}

\title{A stochastic computing architecture for \\ iterative estimation}

\author{Michael~Lunglmayr,
		Daniel~Wiesinger,
		Werner~Haselmayr
\thanks{Johannes Kepler University Linz, Austria, corresponding e-mails: michael.lunglmayr@jku.at, werner.haselmayr@jku.at}
}

\maketitle

\begin{abstract}
Stochastic computing (SC) is a promising candidate for fault tolerant computing in digital circuits. We present a novel stochastic computing estimation architecture 
allowing to solve a large group of estimation problems including least squares estimation as well as sparse estimation. This allows utilizing the high fault tolerance of stochastic computing for implementing estimation algorithms. The presented architecture is based on 
the recently proposed linearized-Bregman-based Sparse Kaczmarz algorithm. To realize this architecture, we develop a shrink function in stochastic computing and analytically describe its error probability.
We compare the stochastic computing architecture to a fixed-point binary implementation and present bit-true simulation 
results as well as synthesis results demonstrating the feasibility of the proposed architecture for practical implementation.
\end{abstract}


\IEEEpeerreviewmaketitle

\section{Introduction}
Stochastic computing is a promising candidate to increase the fault tolerance of digital circuits. Distributing the weights of a digital number evenly over a long bitstream allows for a high robustness against errors.
Stochastic computing circuits have been successfully used in areas such as decoding of error correcting codes, control systems, image processing, filter design, and neural networks (see e.g.~\cite{Alaghi_17,Alaghi_13} and the references therein).

Estimation algorithms represent an important class of signal processing methods. Their applications range from Radar and communications engineering, over speech and image processing to bio-medical applications \cite{Kay97}. Because estimation algorithms are typically used in applications with noisy measurements, estimation results are naturally afflicted with errors. Due to these inevitable errors, approximate algorithms are often used in this field, allowing for a trade-off between precision and computational complexity. For this reason, approximate estimation algorithms seem to be perfectly fitted to the stochastic computing world, as both are based on similar maxims, putting algorithmic robustness before exact solutions. However, the number of estimation algorithms realized using stochastic computing has been limited so far. 
%
%
%
%
According to our opinion, this is due to several reasons:
\begin{itemize}
\item \emph{Many estimation algorithms are not stream-based and thus not very suited for SC implementation.} Examples are algorithms that require the access to full matrices such as the sequential least squares (SLS) \cite{Kay97} or that use branching operations such as such as active-set based LASSO algorithms for sparse estimation \cite{friedman2007pathwise}. 
\item \emph{Unfavorable properties of some stochastic representation formats for the requirements of estimation algorithms.} When considering e.g. sparse estimation, most elements of an estimation result are zero. Using e.g. the single-line bipolar SC format, zeros are translated into bitstreams with a one-probability of $0.5$. For finite bitstreams this potentially introduces errors (randomly generated bitstreams might not have exactly half of its bits equal to one) and leads to a frequent switching of the computation elements resulting in a high energy consumption. 

\item \emph{Estimation algorithms often require arithmetic operations that have non-beneficial properties 
for SC.} For example, estimation algorithms often require the calculation of scalar products with a 
large number of inputs. When implementing such scalar products by e.g. using a tree of scaled adders this would 
result in very large computation errors for practical bitstream lengths, often rendering the result unusable.
\end{itemize}

In this work, we consider all of these three aspects, and present an architecture realizing a linearized Bregman based Sparse Kaczmarz algorithm allowing performing a large class of estimation algorithms with stochastic computing: combined {$l_1/l_2$-norm estimation}. This class of algorithms allows solving least squares estimation problems as well as sparse estimation problems for estimation scenarios based on a blocks of measurement data, or when used as adaptive filters, it allows performing least mean squares (LMS) filters or Sparse LMS filters \cite{LBLMS}. This allows to cover a large part of the field of estimation in signal processing and data analysis, from traditional problems that are often based on $l_2$-norm estimation to more recent estimation problems such as in compressive sampling \cite{candes2008dict} that can be solved with combined {$l_1/l_2$-norm estimation}.

A realization of an LMS filter using stochastic computing has already been presented in \cite{LMSSC}. There, the authors used a parallel counter in the scalar product followed by a converter for the binary number output by the parallel counter to a stochastic bitstream. Although, this represents an interesting design, it uses binary counters in the core algorithm. Our aim, however, was to develop a fully stochastic design (except for the necessary storage between the iterations), avoiding any conversions to binary numbers in the core algorithm. This e.g. allows ensuring the high fault tolerance of stochastic computing throughout the whole algorithmic core. 

\section{Kaczmarz type algorithms for Estimation}
\label{Sect:Kacz}
The algorithms discussed in this work are based on a linear model of the form
\begin{align}
{\bf y} = {\bf A}{\bf x} + {\bf w}. \label{eqn:sysmodel}
\end{align}
Here, ${\bf y}$ is a measurement vector, ${\bf A}$ a known matrix, often called system matrix, ${\bf w}$
an unknown noise vector and ${\bf x}$ the unknown vector that is to be estimated. The dimensions of all
vectors in (\ref{eqn:sysmodel}) are defined by the dimension of the matrix ${\bf A}$: $m \times n$. 
Algorithms for estimating ${\bf x}$ from the measurements ${\bf y}$ often require matrix-vector operations resulting in a large number of memory access operations (e.g. as in the SLS algorithm). For a stochastic computing implementation, memory access operations should be avoided as much as possible, due to the required conversion effort for bit-stream generation and for conversion  
into the storage format. One would prefer algorithms that work in a stream-based fashion, avoiding excessive use of memory access operations.

A promising type of algorithms allowing performing estimation of ${\bf x}$ are Kaczmarz-type algorithms \cite{Kaczmarz37, ALS}. These algorithms do not require branching operations and use only vector operations avoiding the more costly matrix-vector operations. Traditionally, these algorithms have been used for $l_2$-norm estimation problems using cost functions of the form
\begin{equation}
\underset{{\bf x}}{\text{arg min }} \|{\bf y} - {\bf A}{\bf x} \|_2^2.
\label{eqn:l2cost}
\end{equation}
The Kaczmarz algorithm is used to approximately solve this problem \cite{Kaczmarz37, ALS}.
Recently, an extension based on linearized Bregman iterations has been proposed \cite{LBKaczmarz} allowing performing sparse estimation as well. This is done using the combined $l_1/l_2$ cost function
\begin{equation}
\underset{{\bf x}:\; {\bf A}{\bf x}={\bf b}}{ \text{arg min }} \frac{1}{2}\| {\bf x} \|_2^2 + \lambda \| {\bf x}\|_1.
\label{eqn:l1cost}
\end{equation}
The value $\lambda$ is used to control the sparsity of the estimation result. The higher this value, the lower the number of non-zero values in the estimation result will be. 
Based on so-called linearized Bregman iterations \cite{Osher4, Yin, Yinlin} this allows to formulate the following Sparse Kaczmarz algorithm \cite{LBKaczmarz}, shown in Alg.~\ref{alg:Kaczmarz} for solving $(\ref{eqn:l1cost})$.
\vspace{-.3cm}
\begin{algorithm}[h]
\caption{Sparse Kaczmarz (SK)}
\label{alg:Kaczmarz} 
\small
\begin{algorithmic}[1]
\Require ${\bf y}$, ${\bf A}$, $\lambda$
\Ensure $\hat{\bf x}^{(N)}$
\State{$\hat{\bf x}^{\left(0\right)} \leftarrow {\bf 0} $}
\State{$\hat{\bf v}^{\left(0\right)} \leftarrow {\bf 0}$}
\For{$k=1..N$}
\For{$j=1..n$}
\State{$\hat{ x}_j^{\left(k\right)}\leftarrow $ shrink $\left(\hat{v}_j^{\left(k\right)},\lambda\right)$}
\EndFor
\State $i \leftarrow ( (k - 1) \text{ mod }m ) + 1$ \Comment {cyclic re-use of rows of ${\bf A}$}
\State{$\hat{\bf v}^{\left(k+1\right)}\leftarrow \hat{\bf v}^{\left(k\right)}+\frac{1}{\|{{\bf a}_i}\|_2^2}{{\bf a}_i}\left(y_i-{{\bf a}_i}^T\hat{\bf x}^{\left(k\right)}\right)$}
\EndFor
\end{algorithmic}
\end{algorithm}
\vspace{-.25cm}

\noindent
Here, ${\hat{\bf x}^{(k)} = (\hat{x}_1^{(k)}, \ldots, \hat{x}_n^{(k)})^T}$ and 
${\hat{\bf v}^{(k)} = (\hat{v}_1^{(k)}, \ldots, \hat{v}_n^{(k)})^T}$ for $k = 1..N$, with $N$ as the overall number of iterations. ${\bf a}_i$ are the rows of ${\bf A}$.
The shrink function of Alg.~\ref{alg:Kaczmarz} is defined as 
\begin{align}
\label{equ:shrink}
	 \text{shrink}(v, \lambda) = \text{max}(\vert v\vert - \lambda, 0)\text{sign}(v).
\end{align}
As it is shown in \cite{ISCAS2017}, by an appropriate scaling of $\lambda$ as well as the measurements ${\bf y}$, one can perform the whole algorithm in fractional precision fixed-point using only values out of the interval $[-1,1)$ (for arbitrary lambda values that are then scaled down as well). Assuming one wants to perform an estimation with a value $\lambda'$, one typically sets the algorithm's $\lambda$ value to the fixed value of $0.5$ and scales 
the measurement values by $0.5/\lambda'$. The algorithm will then automatically output the scaled estimation result $\hat{\bf x}^{(N)}\cdot 0.5/\lambda'$ \cite{ISCAS2017}. This scaling method using $\lambda=0.5$ in the algorithm was also utilized for the sparse estimations of this work.

If one sets the parameter $\lambda$ to zero, the output of shrink
is the same as its input ${\text{shrink}(v, 0) = v}$. In this case, Alg. $1$ becomes the ordinary Kaczmarz algorithm (for the estimation problem of (\ref{eqn:l2cost}) ) using ${ {\bf x}}^{\left(k\right)} = { {\bf v}}^{\left(k\right)}$. 
As one can notice from Alg. $1$, Kaczmarz algorithms re-use measurement values as well as rows from the matrix ${\bf A}$, in a cyclic way. This is
ensured by the modulo operation in line $7$ of the algorithm. However, if one assumes ${\bf A}$ to be a convolution matrix and uses $N=m$, Alg.~$1$
performs the operations of a normalized LMS (NLMS) \cite{haykin}. If one furthermore uses a $\lambda$ value larger than zero, the algorithm becomes the Sparse LMS from \cite{LBLMS}. Tab.~\ref{tab:mani} shows the manifoldness of the basic algorithm. 

\begin{table}
\caption{ Manifoldness of Algorithm 1 \label{tab:mani}}
\begin{center}
\begin{tabular}{|c||c|c|}
\hline
Alg. 1 becomes & $\lambda = 0$ & $\lambda > 0$\\ 
\hline
\hline
arbitrary ${\bf A}$, $N > m$ & Kaczmarz & Sparse Kaczmarz \\
\hline
${\bf A}$ convolution matrix, $N = m$ & NLMS & Sparse LMS \\
\hline
\end{tabular}
\end{center}
\vspace{-.75cm}
\end{table}
In addition to the versatility of the algorithm, its simple algorithmic structure makes the algorithm perfectly suited to be implemented in stochastic computing. Using no branching operations and only a small amount of storage operations fosters the stochastic computing implementation described below. 
However, for an efficient SC realization, an appropriate stochastic computing format has to be chosen and an efficient implementation of the non-trivial shrink function is required. These aspects are discussed in the next section. 
\section{Stochastic implementation of \\Kaczmarz type algorithms} 
\subsection{Stochastic Computing formats}
For stochastic computing, often single-line formats such as the unipolar or the  bipolar format \cite{Gaines_69} are used. 
Choosing a certain stochastic computing format naturally affects how arithmetic operations are performed. For the unipolar format, huge variaties of arithmetic operations have been designed (see e.g. \cite{Alaghi_13} and the references therein).
Using a unipolar format only allows for non-negative numbers. This is often resolved by using the single-line bipolar format. However, compared to the unipolar format this format changes and sometimes even complicates the arithmetic operations. 
As an example, to the best of the authors' knowledge, there does not exist an architecture for a non-scaled adder for this format. This lack of a practical architecture for a non-scaled adder prevents using this format for many signal processing applications, especially 
for estimation purposes. 

The drawbacks of single-line formats can be relieved when introducing a second line. 
One representation is the so-called signed magnitude format \cite{Toral_00}. It represents a deterministic number ${x \in [-1,1]}$ by two stochastic bitstreams, a sign stream ${X}_\text{s}$ and a magnitude stream ${X}_\text{m}$. While the magnitude stream can be interpreted as a number in the unipolar format, the sign stream represents the corresponding sign of a bit in the magnitude stream.  The value of a number is defined as
\begin{align}
  x & = \frac{1}{L} \sum\limits_{l=1}^{L} \left(1-2X_\text{s}[l]\right) X_\text{m}[l] ,
  \label{eq:sgn_magn_conv_repr}
\end{align}
where $L$ is the length of the bitstreams, and $X_\text{s}[l]$ and $X_\text{m}[l]$ are the $l^{th}$ bits of the sign and magnitude streams, respectively.
%

\begin{figure}[h]
	\centering
	\includegraphics[width=0.25\columnwidth]{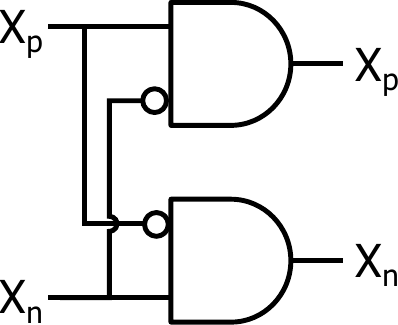}
	\caption{Cancellation circuit eliminating a $1$ in $X_\text{n}$ and $X_\text{n}$ at the same time.}
	\label{fig:cancellation}
\end{figure}

Similar to the signed magnitude format, the two-line bipolar format (TLB) \cite{Gaines_69} represents a deterministic number ${x \in [-1,1]}$ by two unipolar stochastic bitstreams, the positive stream ${X}_\text{p}$ (its ones are counted positive) and 
the negative stream ${X}_\text{n}$ (its ones are counted negative). 
A value in the TLB format is defined as
\begin{align}
  x & = \frac{1}{L}\sum\limits_{l=1}^{L} X_\text{p}[l] - X_\text{n}[l],
  \label{eq:tlb_conv_repr}
\end{align}
where $X_\text{p}[l]$ and $X_\text{n}[l]$ are the $l^{th}$ bits of the positive and negative streams, respectively.
This two-line representation efficiently allows using shift register based arithmetic units enabling a high precision implementation of e.g. a SC scalar product \cite{SCScalar}. Furthermore, the signed magnitude and the two-line bipolar format can be easily converted into each other using only two logic gates \cite{SCScalar}. 
Fig.~\ref{fig:cancellation} shows a cancellation circuit, described in \cite{Gaines_69, SCScalar} allowing a minimum variance representation \cite{Gaines_69} of a TLB represented number at its output. It simply works by cancelling two ones appearing in the same position in the $p$ and the $n$ stream.
The TLB representation enables an efficient implementation of the shrink function for the Sparse Kaczmarz algorithm, as we describe below.

\subsection{Stochastic computing shrink function}

The implementation of the Sparse Kaczmarz algorithm requires the following building blocks: non-scaled adders, multipliers, a scalar product and shrink function blocks. Except for the shrink function, all 
of these building blocks have been already described in the literature. We used the non-scaled adders, multipliers as well as the sequential-shift scalar product from \cite{SCScalar} in our implementation of the Sparse Kaczmarz algorithm.

In the following, we describe the novel design of the shrink function in stochastic computing. 
Shrink is a non-smooth function. It sets its output to zero if the absolute value of its input is smaller than $\lambda$ and else reduces the magnitude of the 
input by $\lambda$. Although, at a first glance, it might seen that this would 
require a branching operation, it can be performed in a stream-based manner as well.

The shrink operation will be done right after the iteration update. For this, the stochastic  bitstreams are first converted to a storage representation, e.g. as described in \cite{SCStore}. Then new bitstreams are then re-generated for the next iteration. We assume that the new bitstreams are generated such that one of the two lines is all-zero. This can be e.g. performed by using Memristors. As the authors of \cite{SCStore} describe, with Memristors one can convert an analog voltage to a stochastic bitstream and vice versa. For a two line format, one could subtract the voltages from the $p$ and $n$ streams in the analog domain and then generate the corresponding bitstream out of the difference voltage. 

\begin{figure}[bh]
	\centering
	\includegraphics[width=\columnwidth]{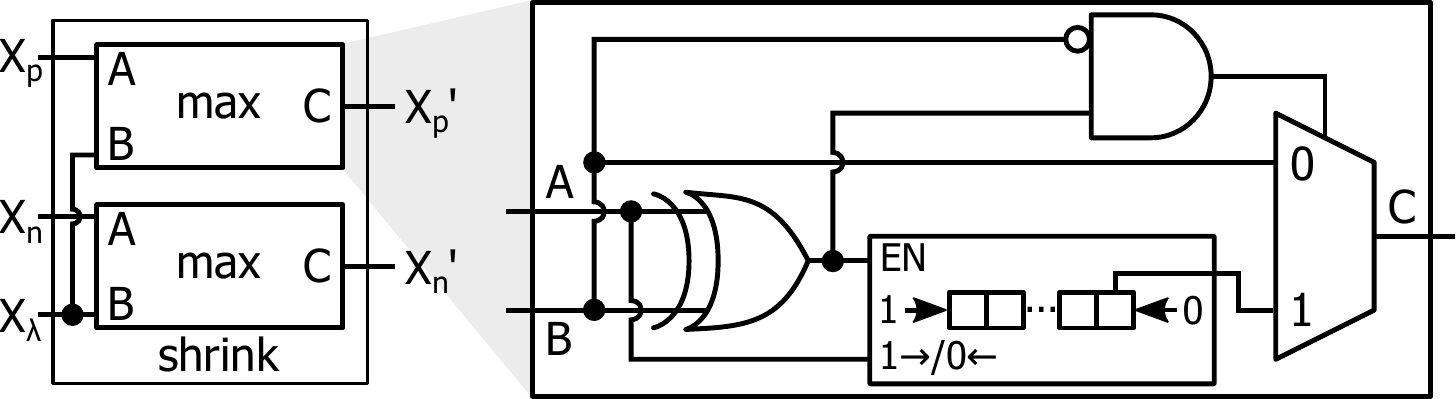}
	\caption{Stochastic computing implementation of the shrink function.}
	\label{fig:scshrink}
\end{figure}

Using this method of generation of the TLB stream allows for an efficient implementation of the shrink operation, as we describe below. Fig.~\ref{fig:scshrink} shows the stochastic computing implementation of the shrink function using two stochastic maximum circuits. 

In the literature, several min/max circuits for stochastic computing have been proposed \cite{Li_11, Yu_17, NovelMaxMin}. The approach of \cite{NovelMaxMin}, as depicted on the right in Fig.~\ref{fig:scshrink}, not only provides a higher precision compared to the other proposed circuits, but also shows an important property for implementing the shrink function:
the ones of input stream $B$ are always output by the circuit. This can be easily confirmed by analyzing the circuit of Fig.~\ref{fig:scshrink}.
The circuit has a bi-directional shift register to prevent ones of the input stream $A$ to be output if the stream of $A$ represents a smaller
number than $B$. If $A$ represents a larger number than $B$, ones of $A$ in excess to $B$ are output as well, together representing the number of bitstream $A$ (then the maximum) in the output stream. As described in \cite{NovelMaxMin}, 
this circuit closely approximates the maximum function with high precision, with its error depending on the shift register length as well as of course on the bitstream length.

The property that the ones of $B$ are always output by the circuit enables an efficient implementation of the shrink function. Assuming large enough shift registers, by connecting the bitstream $X_\lambda$ (e.g. for the typical value $\lambda=0.5$) to both $B$ inputs of the two stochastic maximum blocks, both blocks should output the $X_\lambda$ stream if the magnitude of the input value is smaller than lambda (one of the numbers represented by $X_\text{p}$ or $X_\text{n}$ then being smaller than $\lambda$ and the other one being zero). 
If one of the values represented by the inputs $X_\text{p}$ or $X_\text{n}$ is larger than $\lambda$, meaning that the magnitude of the TLB-represented value is larger than $\lambda$, one output of the stochastic maximum blocks will represent this larger value, the other one (due to having a zero bitstream at its input $A$) will output the stream $X_\lambda$. Then, the TLB representation will naturally provide the subtraction of the values represented by $X_\text{p}$ and $X_\text{n}$. 
For the case of both maximum circuits outputting the stream $X_\lambda$, a following cancellation block from Fig.~\ref{fig:cancellation} will produce two all-zero streams.

When using this stochastic shrink function, especially the error for cases when the shrink function should output a zero value is most relevant. This is because in sparse estimation it is most relevant which elements of the estimated vector are non-zero, thus an error at zero positions (leading to a new non-zero element) typically has a more severe effect than an error at non-zero positions.
For the maximum block with 
the all-zero stream at input $A$, the output will always be exactly $X_\lambda$. For the other block, the output might deviate from $X_\lambda$. 
The probability of an error of a bit in this output stream can be analytically described\footnote{The error for the SC shrink function when the output should not represent the zero value can be analytically described as well but is omitted due to its lower relevance and page restrictions}.
Due to the implicit subtraction of the TLB format, the bit error probability of this second stream deviating from 
$X_\lambda$ is equal to the bit error probability of 
shrink's output compared to the optimal zero streams. We will call this error probability $P_{!0}$. 

For the stochastic maximum, an analytic description of the error probability of the output stream for the case that the value of bitstream of input $A$ is smaller than of $B$ is given as \cite{NovelMaxMin}: 
\begin{align}
  P_{e, a \leq b} = \left( \frac{\left(\frac{P_A(1-P_B)}{P_B(1-P_A)}\right)^{M}}{\sum\limits_{j=0}^{M} \left(\frac{P_A(1-P_B)}{P_B(1-P_A)}\right)^j} \right )  P_A (1-P_B),
\label{eqn:Peab}
\end{align}
with $M$ as the length of the shift register. $P_A$ and $P_B$ are the probabilities of 
having bit one in the bitstreams of $A$ and $B$, respectively, and thus specify the (unipolar) values of these bitstreams.

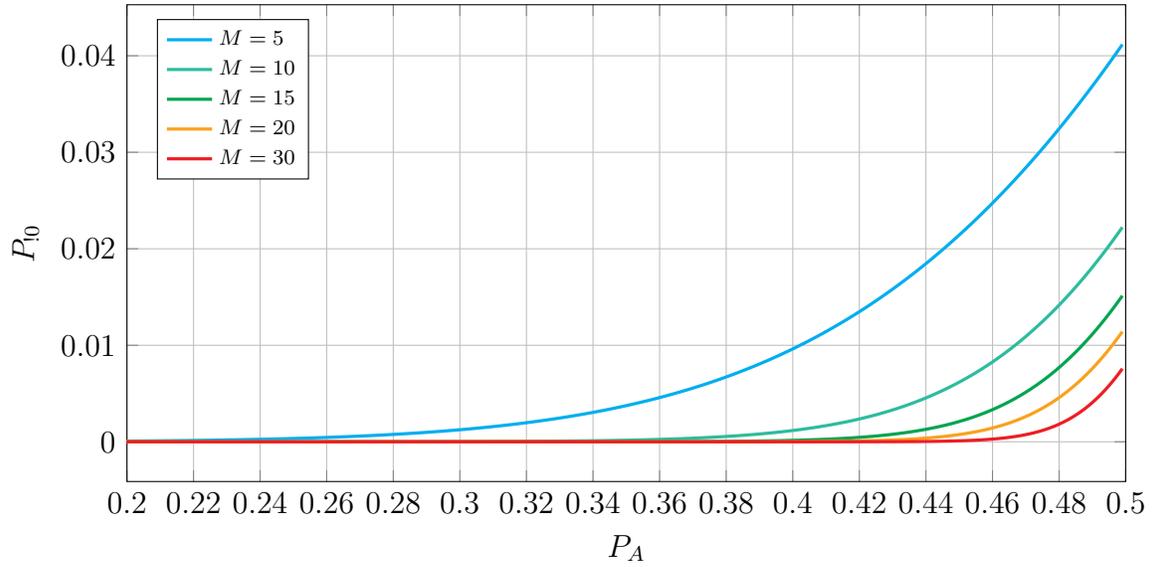
\begin{figure}[t]
\begin{center}
\vspace{-.03cm}
\begin{tikzpicture}
\begin{axis}[compat=newest, 
width=.9\columnwidth, height =.48\columnwidth,log basis y=10, grid,
ylabel={ $P_{!0}$ }, 
xlabel={ $P_A$ }, 
legend pos=north west, 
xmin = .2,
xmax = .5,
ytick={0,0.01,0.02,0.03,0.04},
legend cell align=left,
scaled y ticks=false,
yticklabel style={
        /pgf/number format/fixed,
        /pgf/number format/precision=2
},
]
\addplot[color=Cyan,very thick] table[x index =0, y index =2] {./PeForL.dat};
\addlegendentry{ \scriptsize $M=5$}
\addplot[color=SeaGreen,very thick] table[x index =0, y index =3] {./PeForL.dat};
\addlegendentry{ \scriptsize $M=10$}
\addplot[color=Green,very thick] table[x index =0, y index =4] {./PeForL.dat};
\addlegendentry{ \scriptsize $M=15$}
\addplot[color=YellowOrange,very thick] table[x index =0, y index =5] {./PeForL.dat};
\addlegendentry{ \scriptsize $M=20$}
\addplot[color=Red,very thick] table[x index =0, y index =6] {./PeForL.dat};
\addlegendentry{ \scriptsize $M=30$}

\end{axis}
\end{tikzpicture}
\caption{ Non-zero error probabilities of shrink for different shift register lengths. \label{fig:PeShrOverL}}
\end{center}
\vspace{-.7cm}
\end{figure}

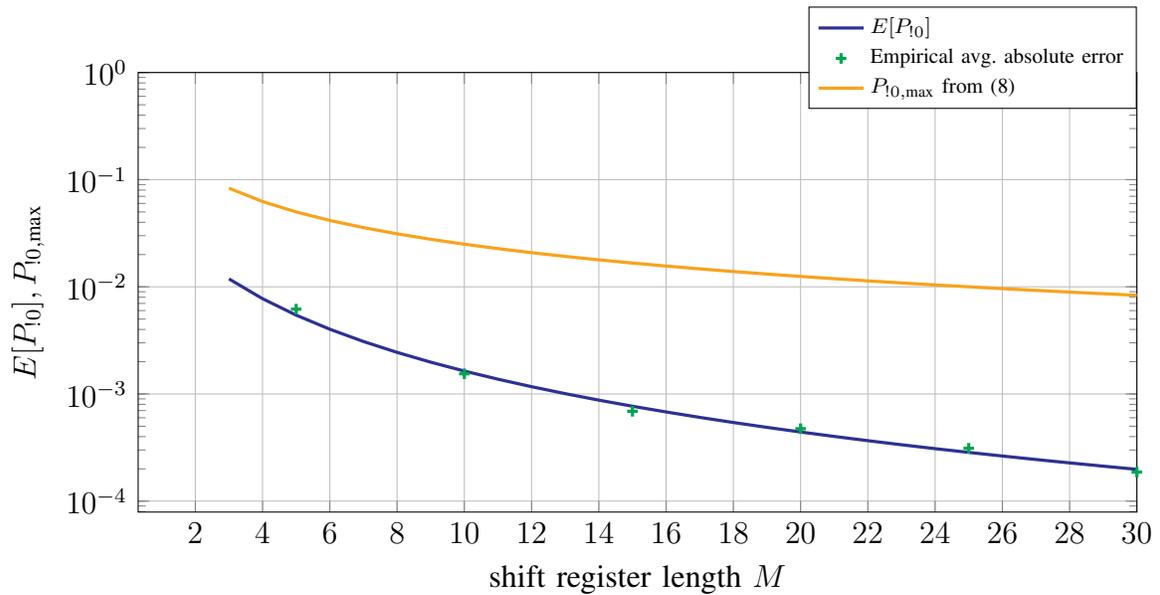
\begin{figure}[t]
\begin{center}
\vspace{-.03cm}
\begin{tikzpicture}
\begin{semilogyaxis}[compat=newest, 
width=.9\columnwidth, height =.45\columnwidth,log basis y=10, grid,
ylabel={ $E[ P_{!0} ], P_{!0,\text{max}}$ }, 
xlabel={ shift register length $M$ }, 
legend pos=north east, 
ymax = 1,
xmax = 30,
legend style={at={(1.0,1.15)}},
legend cell align=left,
]
\addplot[color=Blue,very thick] table[x index =0, y index =1] {./MaxAndExpectedPe.dat};
\addlegendentry{ \scriptsize $E[ P_{!0} ]$}
\addplot[color=Green,very thick, only marks, mark=+] table[x index =0, y index =1] {./Empirical.dat};
\addlegendentry{ \scriptsize Empirical avg. absolute error }
\addplot[color=YellowOrange,very thick] table[x index =0, y index =2] {./MaxAndExpectedPe.dat};
\addlegendentry{ \scriptsize  $P_{!0,\text{max}}$ from (\ref{eqn:e_upper_bound})}
\end{semilogyaxis}
\end{tikzpicture}
\caption{ Expected and maximum error probabilities of shrink. The empirical avg. absolute error was obtained by averaging over $1000$simulations for each value $M$ using $L=10^6$.\label{fig:expshrinkerror}}
\end{center}
\vspace{-.5cm}
\end{figure}

For use in the shrink function, this equation can be simplified by setting $P_B = 0.5$ (the typically used $\lambda$ value as described in Sec.~\ref{Sect:Kacz}), resulting in $P_{!0} = P_{e, a \leq 0.5}$. Fig.~\ref{fig:PeShrOverL} shows 
the evaluation of (\ref{eqn:Peab}) for values $P_A$ up to $0.5$ using different values $M$. As one can see, the error probabilities always have their maximum for $P_A$ values of $0.5$. This maximum error probability can be used as upper bound of the actual error probability for all values $P_A$. Via the limit $P_A \rightarrow 0.5$ for $P_B = 0.5$ in (\ref{eqn:Peab}), one can analytically find this maximum as 
\begin{equation}
P_{!0,\text{max}} = P_{e, a \rightarrow 0.5 \leq 0.5} = \frac{1}{M} 0.5^2.
\label{eqn:e_upper_bound}
\end{equation}
The expected value of the error probability can be 
found by integrating $P_{!0} = P_{e, a \leq 0.5}$
over all values of $a$ (assuming a uniform distribution of the values of $a$). 
To the best of our knowledge, this integration can only be performed numerically. Fig.~\ref{fig:expshrinkerror} shows the expected error probabilities
as well as the maximum value (\ref{eqn:e_upper_bound}) for different shift register lengths $M$. We also plotted empirical results validating the 
theoretical results. The simulation times to obtain the 
empirical results was approximately a factor $4\cdot10^4$ larger than 
the time required for numerical integration. One can see from these figures, that a good error performance can be already achieved for moderate shift register lengths $M$.

%
%

\subsection{Stochastic Computing Sparse Kaczmarz Architecture}
\begin{figure}[h]
	\centering
	\includegraphics[width=.85\columnwidth]{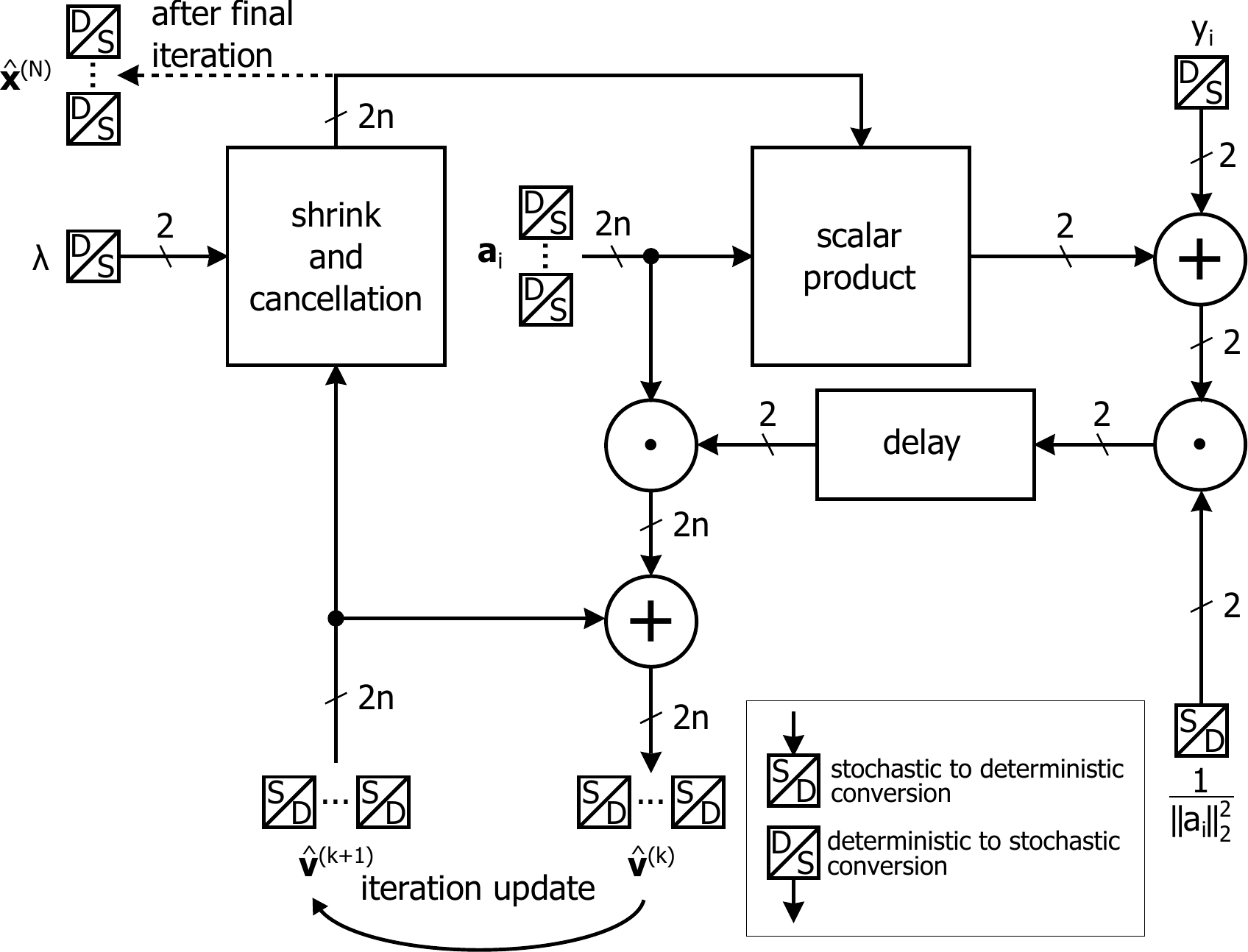}
	\caption{Stochastic computing Sparse Kaczmarz architecture.
	\label{fig:ssk_overview}}
\end{figure}

Combining the building blocks described above, we developed the SC architecture depicted in 
Fig.~\ref{fig:ssk_overview}. As described above, this architecture can also be used to 
realize the ordinary Kaczmarz algorithm when using $\lambda=0$, i.e. by bypassing the shrink function. Using this structure in an adaptive filter setting, e.g. assuming that the vector ${\bf a}_i$ has the structure of the row 
of a convolution matrix, one can realize an LMS or a Sparse LMS with this structure.

The operation of the architecture is as follows. 
Conversion blocks between the (deterministic) memory and the stochastic domain generate the bitstreams
for the computation. When starting the iterations, $\hat{{\bf x}}^{(0)}$ and $\hat{\bf v}^{(0)}$ will be both zero, so their corresponding stochastic bitstreams will be all-zero streams. 

An iteration starts by passing the $n$ two-line streams of $\hat{\bf v}^{(k)}$ through the shrink and cancellation block. This block consists of $n$ shrink blocks in parallel, as described above, each followed 
by a cancellation circuit shown in Fig.~\ref{fig:cancellation}. 

The output of the shrink and cancellation blocks are used in the scalar product with the bitstreams of the matrix row ${\bf a}_i$. After the scalar product is performed, only scalar (i.e. single two-line) operations are performed until the bitstreams representing $y_i-{\bf a}_i^T{\bf x}^{(k)}$ are multiplied by the bitstreams of ${\bf a}_i$. To prevent correlations between the bitstream (the streams of ${\bf a}_i$ are used twice) we included a delay consisting of a $10$ flip-flops shift register. 
After $n$ non-scaled adders, the storage elements of $\hat{\bf v}^{(k)}$ are updated for the next iteration. After all $N$ iterations have been performed, 
the output streams after the shrink and cancellation blocks are converted 
to the memory as $\hat{\bf x}^{(N)}$. Please note that the stream-based architecture prevents the need for storing the values of $\hat{\bf x}^{(k)}$
during the iterations, as it would have to be done in a deterministic implementation.

\section{Performance Results}

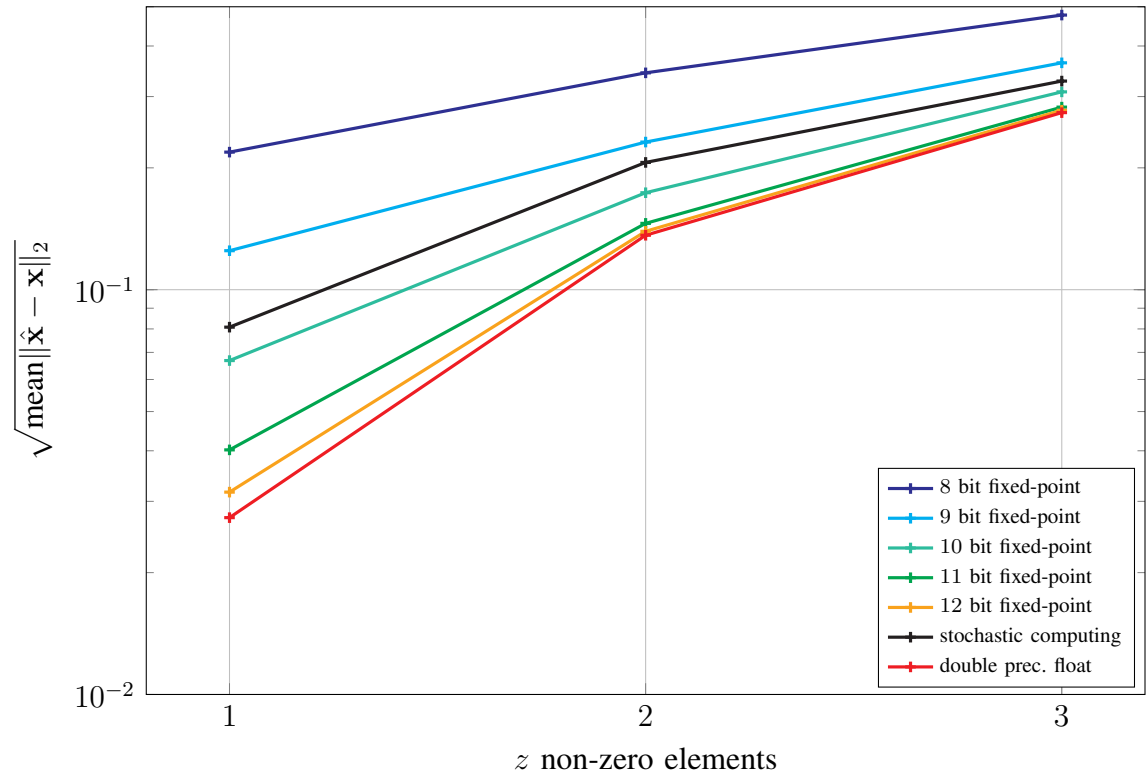
\begin{figure}[t]
\begin{center}
\vspace{-.03cm}
\begin{tikzpicture}
\begin{semilogyaxis}[compat=newest, 
width=.9\columnwidth, height =.65\columnwidth,log basis y=10, grid,
ylabel={ $\sqrt{\text{mean}\|\hat{\bf x}-{\bf x}\|_2}$ }, 
xlabel={ $z$ non-zero elements }, 
ymin = 1e-2,
legend style={at={(0.99,0.01)},anchor=south east},
xtick={1,2,3},
ymax = .5,
legend cell align=left,
]
\addplot[color=Blue,very thick, mark=+] table[x index =0, y index =3] {./error_diagram.dat};
\addlegendentry{ \scriptsize $8$ bit fixed-point }
\addplot[color=Cyan,very thick, mark=+] table[x index =0, y index =4] {./error_diagram.dat};
\addlegendentry{ \scriptsize $9$ bit fixed-point}
\addplot[color=SeaGreen,very thick, mark=+] table[x index =0, y index =5] {./error_diagram.dat};
\addlegendentry{ \scriptsize $10$ bit fixed-point}
\addplot[color=Green,very thick, mark=+] table[x index =0, y index =6] {./error_diagram.dat};
\addlegendentry{ \scriptsize $11$ bit fixed-point}
\addplot[color=YellowOrange,very thick, mark=+] table[x index =0, y index =7] {./error_diagram.dat};
\addlegendentry{ \scriptsize $12$ bit fixed-point}
\addplot[color=Black,very thick, mark=+] table[x index =0, y index =2] {./error_diagram.dat};
\addlegendentry{ \scriptsize stochastic computing }
\addplot[color=Red,very thick, mark=+] table[x index =0, y index =1] {./error_diagram.dat};
\addlegendentry{ \scriptsize double prec. float}
\end{semilogyaxis}
\end{tikzpicture}
\caption{ RMSE for Sparse Kaczmarz. \label{fig:ErrorDiagram}}
\end{center}
\vspace{-.5cm}
\end{figure}

In this section, we show the performance results for the presented SK architecture. Fig.~\ref{fig:ErrorDiagram} shows hardware validated bit-true simulations 
of the stochastic computing implementation described above as well as of the fixed-point design 
from \cite{ISCAS2017} for different bit-widths. The stochastic computing design was implemented for a compressive sampling example estimating ${\bf x}$ vectors with $n=16$ with $z$ non-zero elements at random positions and randomly selected out of $[-1,1]$. The algorithms used $m=10$ measurements per estimation test case. The root mean squared errors (RMSE): $\sqrt{\text{mean} \|{\bf x} - \hat{\bf x} \|_2^2}$ have been averaged over $1000$ simulated test cases. The measurements are modeled using ${\bf A}{\bf x}+{\bf w}$, with the entries of ${\bf A}$ selected uniformly at random from $[-1,1]$ and ${\bf w}$ as white Gaussian 
noise, scaled to obtain a signal to noise power ratio (SNR) of $30$dB.
The bitstreams for the test cases have been generated via maximum length linear feedback shift registers (LFSR) with $L=2^{16} - 2$ (one cycle less than a full period). The scalar product as well as the non-scaled adder (both taken from \cite{SCScalar}) of the design in Fig.~\ref{fig:ssk_overview} use two times shift registers of length $20$ (one for positive and one for negative carries), respectively. 
Based on the evaluations above, we used a shift register length of $10$ as a trade-off between complexity and precision for the SC maximum blocks in the shrink functions. As one can see from Fig.~\ref{fig:ErrorDiagram}, the SC error performance in terms of root mean square errors is in between the performance of the fixed-point implementations of $9$ and $10$ bit respectively. When comparing the synthesis results of the stochastic implementation shown in Tab.~\ref{tab:synth}, one can see that the stochastic computing implementation requires about the same number of combinatorial functions and flip-flops, than the corresponding fixed-point binary implementation of \cite{ISCAS2017}, but require no binary multipliers. 
However, due to the bitstream representation, the stochastic computing requires significantly more clock cycles than the fixed-point binary implementation. On the other hand, as it is exemplary described for the scalar product in \cite{SCScalar}, the robustness in terms of calculation errors of the stochastic computing implementation is significantly higher than the error tolerance of the binary implementation.

\section{Conclusion}
In this work, we presented a fully stochastic computing architecture for performing iterative estimation based on linearized-Bregman-based Sparse Kaczmarz. 
In order to realize this estimation algorithm, we proposed a novel stochastic implementation of the non-linear shrink function and analytically characterized its error performance.
We presented bit true simulation results as well as synthesis results comparing the stochastic computing implementation to a fixed-point binary implementation demonstrating the feasibility of the stochastic computing estimation architecture for practical implementation.

\begin{table}[t]
\caption{ Synthesis Results \label{tab:synth}}
\begin{center}
\begin{tabular}{|l||c|c|}
\hline
Altera Cyclone IV EP4CE115		& SC & fixed-point binary \\
\hline
\hline
Total combinatorial functions   		& $388/114480 $ 			       & $395/114480 $ 			\\
Flip-flops 	& $169/114480 $ 			       & $165/114480 $ 			\\
Embedded 9-bit Multipliers & $0/532 $    & $6/532  $ \\
Fmax Slow 1200mV 85C Model 		& $105.15$ MHz                                        & $102.85$ MHz                              \\       
\hline
\end{tabular}
\end{center}
\vspace{-.5cm}
\end{table}
\ifCLASSOPTIONcaptionsoff
  \newpage
\fi



%

\bibliography{LBSC}

%








\end{document}